\documentclass[twocolumn,superscriptaddress,preprintnumbers,amsmath,amssymb]{revtex4-1}
\usepackage{graphicx}
\usepackage{dcolumn}
\usepackage{bm}
\usepackage{hyperref}
\usepackage{amssymb,amsmath,amsthm,amsfonts}
\usepackage{bbm}
\usepackage{mathtools}
\usepackage{color}
\usepackage{textcomp}
\usepackage{gensymb}
\usepackage{float}
\usepackage{soul}
\usepackage{booktabs}
\usepackage{graphicx}
\usepackage{siunitx}
\usepackage{multirow}
\usepackage{makecell}
\input{Qcircuit}

\usepackage{diagbox}
\usepackage{placeins}

\renewcommand{\sec}[1]{\hyperref[sec:#1]{Section~\ref*{sec:#1}}}
\newcommand{\fig}[1]{\hyperref[fig:#1]{Figure~\ref*{fig:#1}}}
\newcommand{\tab}[1]{\hyperref[tab:#1]{Table~\ref*{tab:#1}}}

\global\long\def\l({\left(}
\global\long\def\r){\right)}

\makeatletter
\newcommand{\vast}{\bBigg@{4}}
\newcommand{\Vast}{\bBigg@{5}}
\newcommand{\VVast}{\bBigg@{11.8}}
\makeatother

\begin{document}


\title{Benchmarking an 11-qubit quantum computer}

\author{K. Wright}
\email{wright@ionq.co}
\affiliation{IonQ, Inc., College Park, MD 20740, USA}
\author{K. M. Beck}
\affiliation{IonQ, Inc., College Park, MD 20740, USA}
\author{S. Debnath}
\affiliation{IonQ, Inc., College Park, MD 20740, USA}
\author{J. M. Amini} 
\affiliation{IonQ, Inc., College Park, MD 20740, USA}
\author{Y. Nam}
\affiliation{IonQ, Inc., College Park, MD 20740, USA}
\author{N. Grzesiak}  
\affiliation{IonQ, Inc., College Park, MD 20740, USA}
\author{J.-S. Chen} 
\affiliation{IonQ, Inc., College Park, MD 20740, USA}
\author{N. C. Pisenti}
\affiliation{IonQ, Inc., College Park, MD 20740, USA}
\author{M. Chmielewski}
\affiliation{IonQ, Inc., College Park, MD 20740, USA}
\affiliation{Joint Quantum Institute and Department of Physics, University of Maryland, College Park, MD 20742, USA}
\author{C. Collins}
\affiliation{IonQ, Inc., College Park, MD 20740, USA}
\author{K. M. Hudek}
\affiliation{IonQ, Inc., College Park, MD 20740, USA}
\author{J. Mizrahi} 
\affiliation{IonQ, Inc., College Park, MD 20740, USA}
\author{J. D. Wong-Campos} 
\affiliation{IonQ, Inc., College Park, MD 20740, USA}
\author{S. Allen}  
\affiliation{IonQ, Inc., College Park, MD 20740, USA}
\author{J. Apisdorf} 
\affiliation{IonQ, Inc., College Park, MD 20740, USA}
\author{P. Solomon} 
\affiliation{IonQ, Inc., College Park, MD 20740, USA}
\author{M. Williams}
\affiliation{IonQ, Inc., College Park, MD 20740, USA}
\author{A. M. Ducore}
\affiliation{IonQ, Inc., College Park, MD 20740, USA}
\author{A. Blinov}
\affiliation{IonQ, Inc., College Park, MD 20740, USA}
\author{S. M. Kreikemeier}
\affiliation{IonQ, Inc., College Park, MD 20740, USA}
\author{V. Chaplin} 
\affiliation{IonQ, Inc., College Park, MD 20740, USA}
\author{M. Keesan}
\affiliation{IonQ, Inc., College Park, MD 20740, USA}
\author{C. Monroe}
\affiliation{IonQ, Inc., College Park, MD 20740, USA}
\affiliation{Joint Quantum Institute and Department of Physics, University of Maryland, College Park, MD 20742, USA}
\author{J. Kim}
\affiliation{IonQ, Inc., College Park, MD 20740, USA}
\affiliation{Department of Electrical and Computer Engineering, Duke University, Durham, NC 27708, USA}

\begin{abstract}
The field of quantum computing has grown from concept to demonstration devices over the past 20 years. Universal quantum computing offers efficiency in approaching problems of scientific and commercial interest, such as factoring large numbers,\cite{Shor1997} searching databases,\cite{Grover1997} simulating intractable models from quantum physics,\cite{Trabesinger2012} and optimizing complex cost functions.\cite{Farhi2014} Here, we present an 11-qubit fully-connected, programmable quantum computer in a trapped ion system composed of 13 $^{171}$Yb$^{+}$ ions. We demonstrate average single-qubit gate fidelities of 99.5$\%$, average two-qubit-gate fidelities of 97.5$\%$, and state preparation and measurement errors of 0.7$\%$. To illustrate the capabilities of this universal platform and provide a basis for comparison with similarly-sized devices, we compile the Bernstein-Vazirani (BV)\cite{Bernstein1997} and Hidden Shift (HS)\cite{vanDam2003,Rotteler2010} algorithms into our native gates and execute them on the hardware with average success rates of 78$\%$ and 35$\%$, respectively. These algorithms serve as excellent benchmarks for any type of quantum hardware, and show that our system outperforms all other currently available hardware. 
\end{abstract}
\maketitle

Small universal quantum computers that can execute textbook quantum circuits exist in both academic\cite{Barends2014, Monz2016, Debnath2016, Roy2018, Erhard2019} and industrial\cite{Chow2012, Devitt2016, Pokharel2018, Hong2019, Nam2019} settings. With a range of two to seventy-two qubits and sufficient fidelity for only tens of entangling gates, these devices and the underlying qubit implementations can be difficult to compare. Even within the trapped ion platform, there is large diversity in atomic species, system architectures, and gate implementations. Trapped ion systems with one to two qubits have shown single-qubit gate fidelities of 99.9999$\%$\cite{Harty2014} with microwave-based operations and better than 99.99$\%$ fidelity with laser based operations,\cite{Gaebler2016,Ballance2016} state preparation and measurement (SPAM) error below $10^{-4}$,\cite{Harty2014,Crain2019} and two-qubit gates with fidelities exceeding 99.9$\%$.\cite{Gaebler2016, Ballance2016} Algorithms have been executed on up to seven trapped-ion qubits\cite{Landsman2018} and, while not optimized for universal
quantum computing, quantum simulators with more than 50 ions have modeled
fundamental quantum systems including Ising chains\cite{Zhang2017} and quantum
magnetism.\cite{Bohnet2016}
\begin{figure}[ht]
\includegraphics[width=\columnwidth]{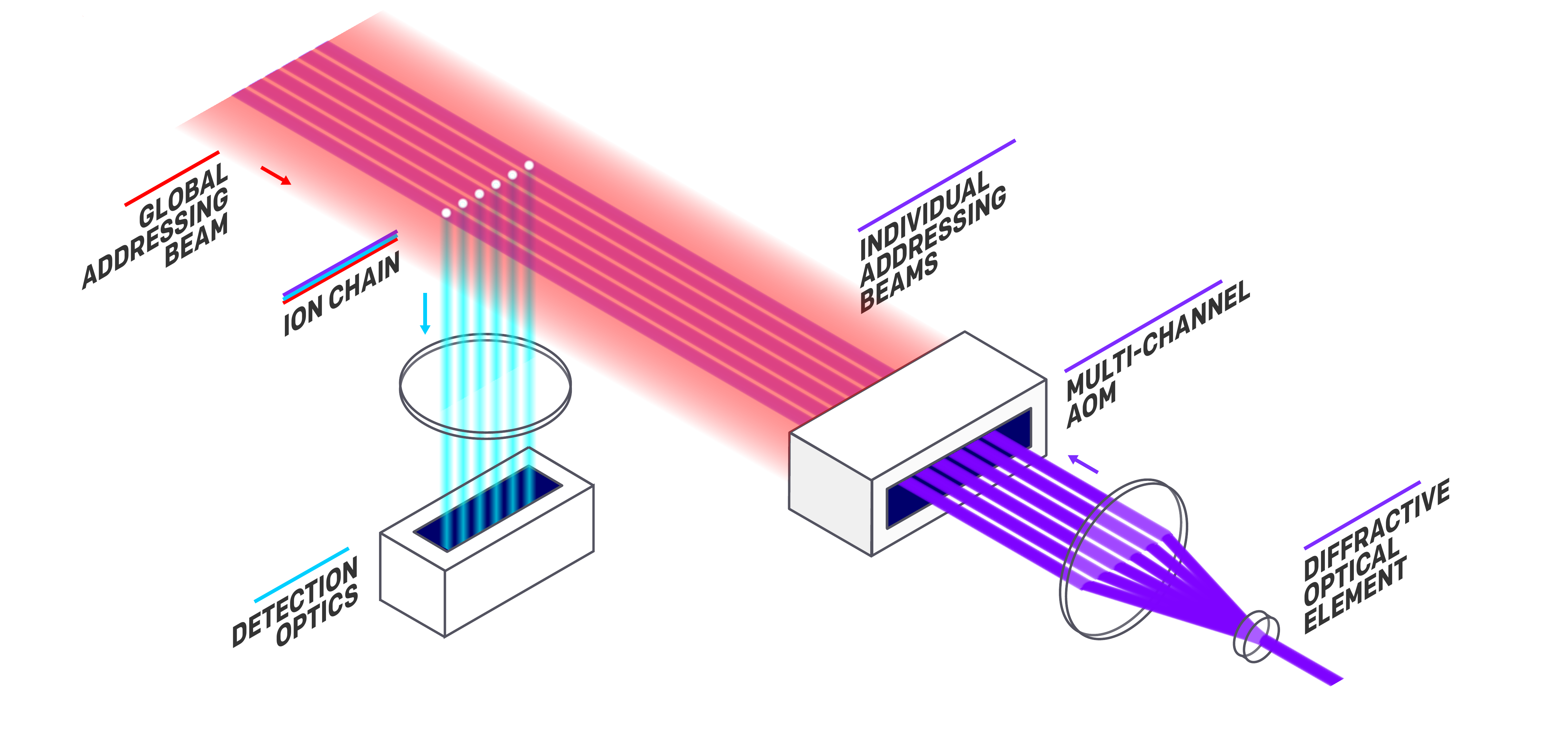}
\caption{Schematic of the hardware. A linear chain of ions is trapped near a surface electrode trap (trap is not shown). Lasers at 369 nm and 935 nm (not shown) illuminate all of the ions during cooling, initialization, and detection. Each ion's fluorescence is imaged through a 0.6 numeric aperture lens (``detection optics'') and directed onto individual photomultiplier tube channels. Two linearly-polarized counterpropagating 355 nm Raman beams are aligned to each qubit-ion, a globally addressing beam that couples to all of the qubits (red) and an individual addressing beam that is focused onto each ion (blue). Acousto-optic modulators (AOMs) modulate the frequency and amplitude of each of these beams to generate single-qubit rotations and XX-gates between arbitrary pairs of qubit ions.}
\label{fig:setup}
\end{figure}

Benchmarking across implementations needs to be both universal across platforms and agnostic to the differences in the underlying hardware. In traditional computing, the performance of computers is measured by executing a set of benchmark problems representing various use-case scenarios, to provide users with an estimate of how the computers would perform in their specific applications. Canonical quantum algorithms demonstrate unambiguous advantage of quantum computers over classical computation, and provide verifiable outcomes to assess successful execution of the algorithm. Therefore, they can serve as ideal candidate problems for benchmarking the performance of any quantum computers. These benchmark algorithms exercise the full hardware/software stack. A hardware-specific compiler breaks down algorithms into the target hardware's native gate set, optimizing for qubit connectivity, gate times, and coherence\cite{Linke2017} to enhance the system's performance.  After execution on the hardware, the measurements can be directly compared with the expected output state to determine the accuracy of the device. This accuracy can then be compared with other devices that have compiled and run the same algorithm.\cite{Murali2019}
\begin{figure}[ht]
\includegraphics[width=\columnwidth]{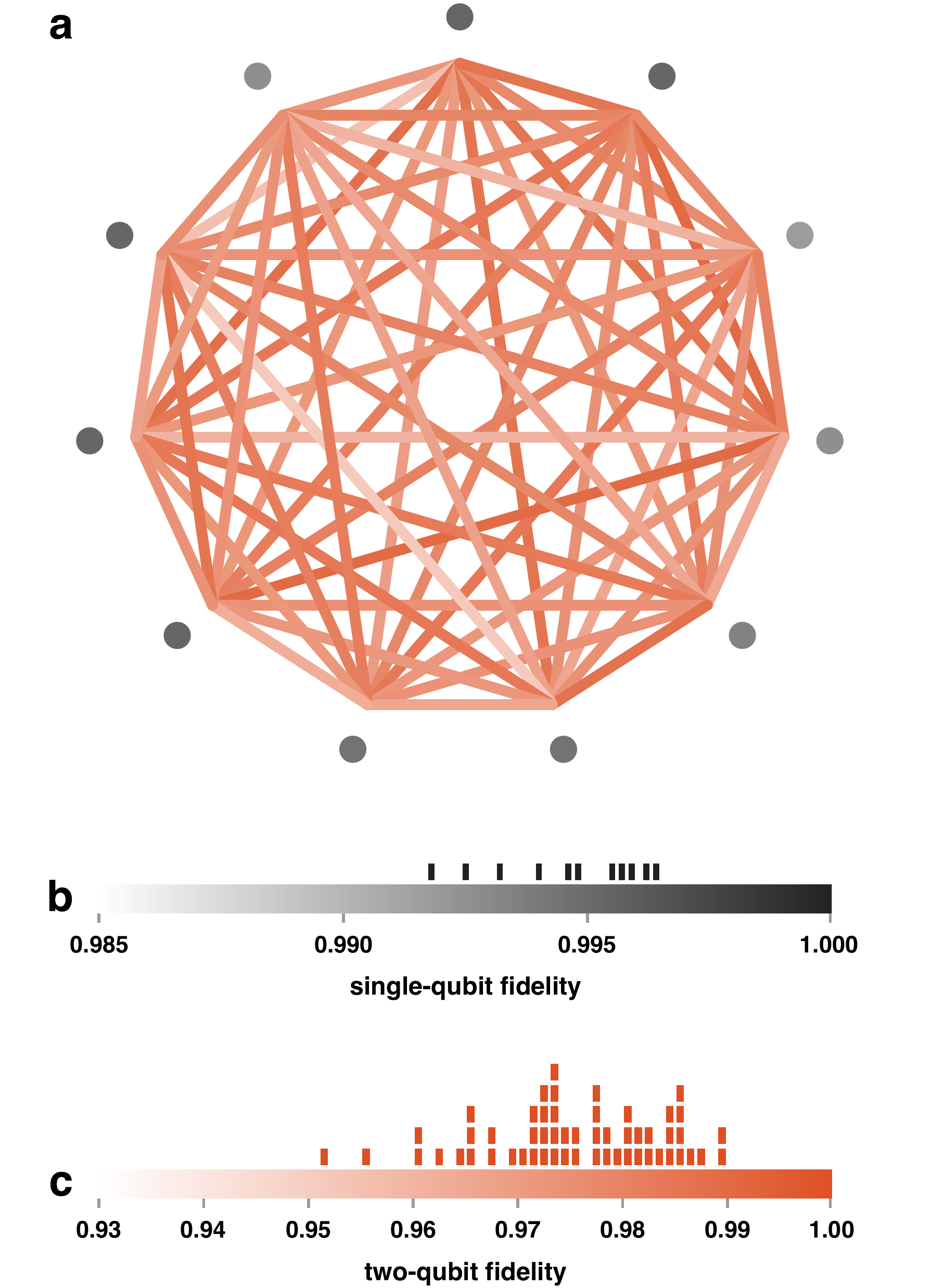}
\caption{Fidelity of native gates. For each qubit pair, we perform an XX-gate and measure the joint populations of the qubit pair as a function of an analysis pulse phase angle. The fidelity of two-qubit gates are plotted as a color scale in the illustration of our all-to-all connectivity in (a). For each qubit, we perform randomized benchmarking to determine the fidelity of the single-qubit gates shown in (b), which are plotted as the color scale of the nodes in (a). We use maximum likelihood estimation to extract fidelities from the parity and joint-population measurement shown in (c). The average two-qubit raw fidelity is 97.5$\%$ and all two-qubit gates perform in the range [95.1$\%$, 98.9$\%$]. The distribution of these fidelities are depicted in the histograms above the color bars shown in (b,c). The fidelity of all single-qubit gates are enumerated in \tab{1Q} and all two-qubit pairs are enumerated in \tab{XX} of the extended data. 
}
\label{fig:XXgates}
\end{figure}

We benchmark two algorithms on an IonQ trapped ion quantum computer, shown schematically in \fig{setup}. Our qubit register is comprised of a chain of trapped $^{171}$Yb$^{+}$ ions, spatially confined near a microfabricated surface electrode trap.\cite{HOA} For this work, we load 13 ions, the middle 11 of which are used as qubits. The two end ions allow for a more uniform spacing of the central 11 ions. The ions are laser-cooled close to their motional ground state using a combination of Doppler and resolved sideband cooling. We encode quantum information into the hyperfine sublevels, $\ket{0}\equiv\ket{F=0, m_F=0}$ and $\ket{1}\equiv\ket{F=1, m_F=0}$ of the $^2S_{1/2}$ ground state. At the beginning of each computation, each qubit is initialized to $\ket{0}$ via optical pumping with high accuracy. After qubit operations (described below), we read out the state of all of the qubits simultaneously by directing laser light resonant with the $^2S_{1/2}$ $\ket{F=1}$ to $^2P_{1/2}$ transition, imaging each ion onto a detector and thresholding the photon counts to determine if each qubit was in the $\ket{1}$ (spin up) or $\ket{0}$ (spin down) state. Thresholding is done by taking a histogram of the collected photons and discriminating between collecting on average zero photons for the  $\ket{0}$ state and ten photons on average for the $\ket{1}$ state.
\begin{figure*}[ht]
\centering
\includegraphics[width=\textwidth]{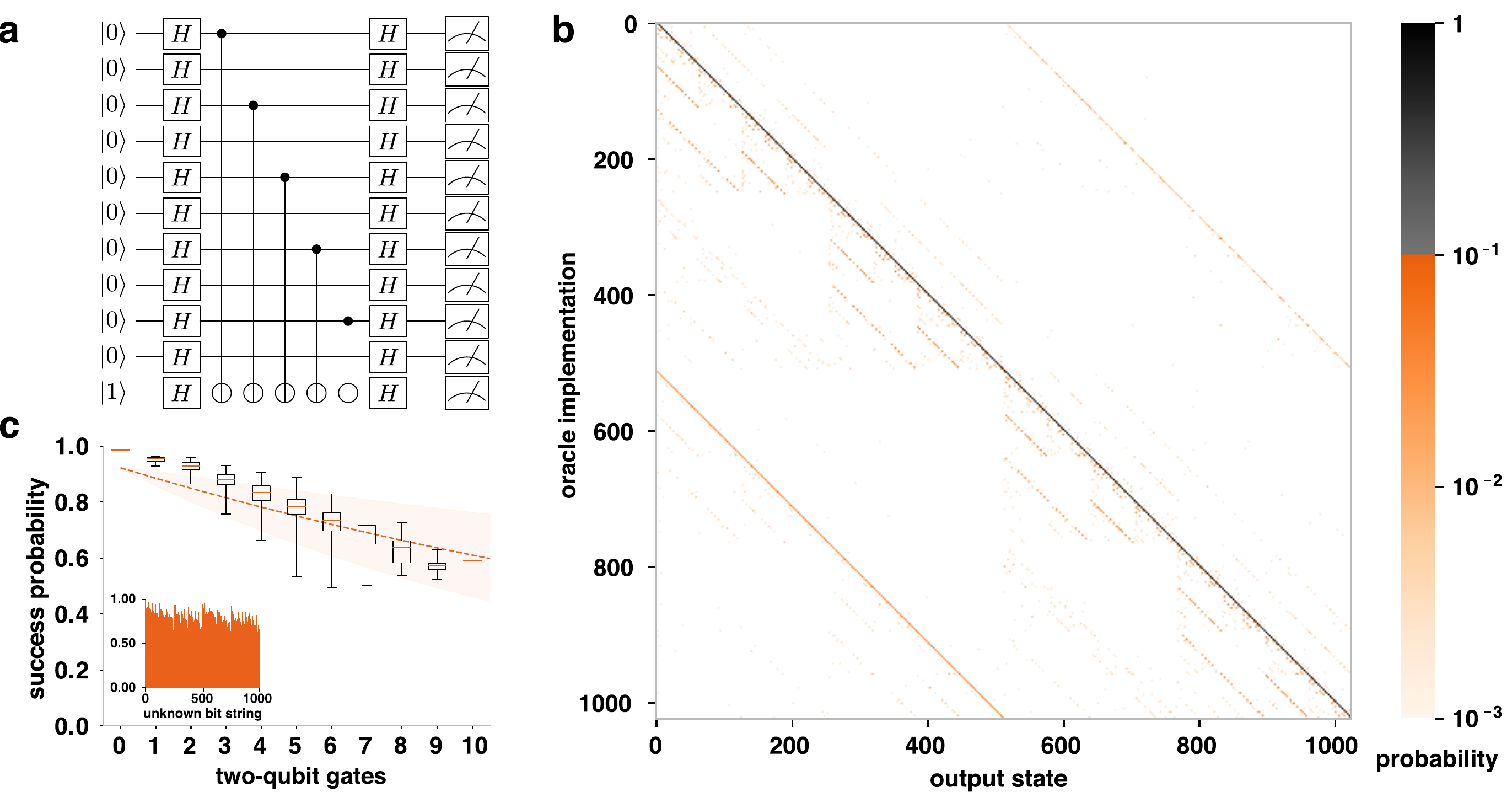}
\caption{Bernstein-Vazirani (BV) algorithm. (a) shows a textbook implementation of the BV algorithm with hidden bit string 1010101010. (b) shows the full output distribution for all 1024 oracle implementations calculated from 500 iterations of each oracle after conditioning on the ancilla. (c) shows the probability (inset plot) of detecting the encoded hidden bit string for all 1024 oracle implementations, as a function of the number of ones in the binary representation of the unknown bit string, which is equivalent to the number of two-qubit gates (n). The boxplots highlight the minimum, first quartile, median, third quartile, and maximum of the data. Note that there is only one oracle implementation for n = 0, 10. The shaded area spans the expected fidelity (excluding crosstalk errors) $\mathcal{F}_\text{2Q}^{\text{n}}\mathcal{F}_\text{1Q}^{2(\text{n}+1)}\mathcal{F}_\text{SPAM}^{10}$ (where $\mathcal{F}_\text{2Q}$ is the fidelity of two-qubit gates, $\mathcal{F}_\text{1Q}$ is the fidelity of single-qubit gates, and $\mathcal{F}_\text{SPAM}$ is the average SPAM fidelity) if all of our gates share the best measured fidelity or, alternatively, all share the worst fidelity. The result of a shared average fidelity is plotted as a dashed line. The average probability of success is 78$\%$ with 899 out of the 1024 oracle implementations exceeding the $2/3$ BQP single-shot success threshold.
}
\label{fig:BV}
\end{figure*}

A two-photon Raman transition drives single- and two-qubit coherent operations by applying a pair of counter-propagating beams from a mode-locked pulsed 355 nm laser.\cite{Hayes2010} One of these beams globally addresses all of the ions simultaneously, while the other beam addresses any of the ions individually (\fig{setup}). The individually-addressing beams pass through a multi-channel acousto-optic modulator (AOM), which allows for the simultaneous modulation of the phase, frequency and amplitude of each beam. To perform a single-qubit gate, we tune the frequency difference between Raman beams to resonantly drive a spin flip transition ($\ket{1}\leftrightarrow\ket{0}$).  In order to perform a two-qubit gate, we off-resonantly drive motional sideband transitions to generate an XX-interaction.\cite{Sorensen2016}  Both the global and individual beams are directed over the trap surface perpendicular to the axis of the ion chain to excite one principal axis of motion transverse to the chain axis. Individual addressing allows us to perform single- and two-qubit gates on any targeted qubits.

Native two-qubit entangling XX-gates are achieved by driving a spin dependent force.\cite{Sorensen2016} Using an amplitude modulated (AM) pulse on any selected pair of qubits, we address multiple transverse motional modes of the ion chain to mediate a spin-spin Ising interaction between qubits.\cite{Choi2014} To achieve high fidelity, the amplitude modulation is calculated to simultaneously decouple all motional modes from the spin at the end of the gate operation. Additionally, these pulse shapes are designed to provide robustness against frequency drift of motional modes and suppress residual off-resonant carrier excitation during the XX-gate.\cite{Choi2014,Wu2018} This gate, in conjunction with single-qubit rotations, forms a universal gate set for performing circuit model quantum computation. Since the XX-gates are mediated by the collective motion of the ion chain, we have all-to-all connectivity between qubits, allowing two-qubit gates to be executed between any qubit pair (\fig{XXgates}a). 

We perform randomized benchmarking\cite{Knill2008} to characterize the single-qubit operations on each ion of the 11-qubit chain. We apply a randomly chosen sequence of $\pi/2$ gates with length $L$ about the $X$ and $Y$ axes. In between each of these $\pi/2$ gates, we either add a $\pi$ rotation about the $X$, $Y$, or $Z$ axis, or an identity operation (leaving the qubit idle for the duration of a gate). A final $\pi/2$ gate is chosen such that the final state is in the $Z$ computational basis (i.e. $\ket{0}$ or $\ket{1}$). We measure the overlap between the measured and expected output states across 500 iterations for at least 24 sequences for each $L \in \{2,4,6,8,10,12\}$. The fidelity of our single-qubit $\pi/2$ gate is then determined by fitting the resulting overlap as a function of sequence length to a power law, $Bp^L+\frac{1}{2}$. Here, the base $p$ is the gate fidelity and the intercept $B+\frac{1}{2}$ is the SPAM fidelity, equivalent to measuring the ion after a single $\pi$ rotation when it is in either state $\ket{0}$ or $\ket{1}$. For a chain of 11 qubits, we measure an average single-qubit fidelity of 99.5$\%$ (\fig{XXgates}b) and an average SPAM fidelity of 99.3$\%$.
\begin{figure*}[t]
\includegraphics[width=2\columnwidth]{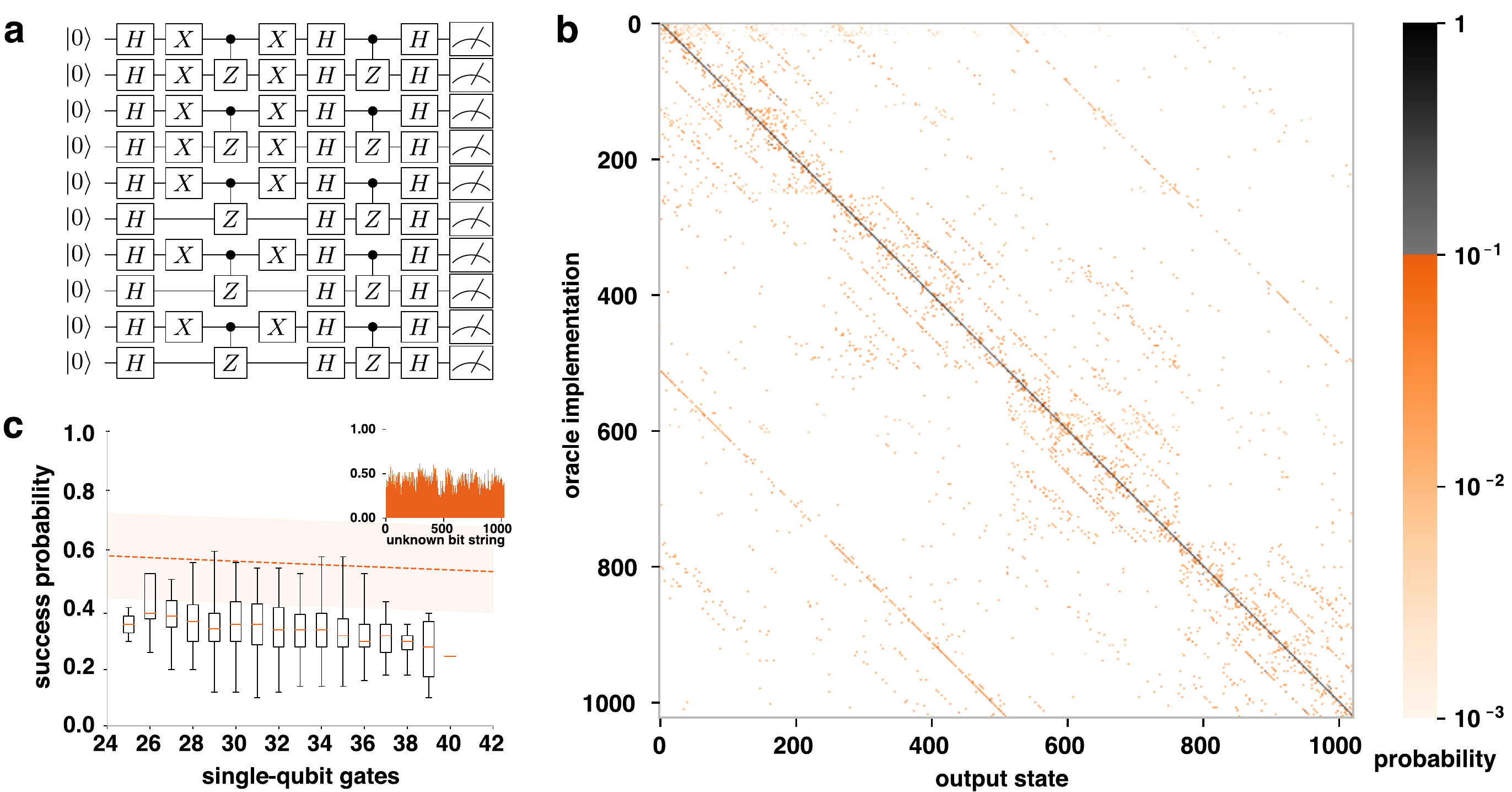}
\caption{Hidden Shift (HS) algorithm implementation on 10 qubits. (a) shows a textbook implementation of the HS algorithm with hidden shift 1111101010. The circuit for each oracle was measured at least 50 times. We trace out the spectator ion and interpret the binary output state of the 10-qubit register as an integer. The full output distribution is shown in (b).(c) shows the probability of detecting the encoded shift s for each of the 1024 oracle implementations versus the number of single-qubit gates (m).  The shaded area represents the expected fidelity $\mathcal{F}_\text{2Q}^{10}\mathcal{F}_\text{1Q}^{\text{m}}\mathcal{F}_\text{SPAM}^{10}$ (where $\mathcal{F}_\text{2Q}$ is the fidelity of two-qubit gates, $\mathcal{F}_\text{1Q}$ is the fidelity of single-qubit gates, and $\mathcal{F}_\text{SPAM}$ is the average SPAM fidelity) if all of our gates share the best measured fidelity or, alternatively, all share the worst fidelity. The result of a shared average fidelity is plotted as a dashed line. The average probability of success is 35$\%$, and 1017 of the 1024 oracle implementations correctly return the hidden shift as the maximal probability state.
}
\label{fig:HS}
\end{figure*}

To quantify the performance of our two-qubit gates and estimate their fidelity, we measure the state fidelity of the Bell state $\frac{1}{\sqrt{2}}(\ket{00}+e^{i\phi}\ket{11})$ prepared using a single XX-gate by performing partial tomography of the state.\cite{Ballance2016, Gaebler2016} The diagonal terms of the two-qubit density matrix are extracted by measuring the populations in the even parity states. The populations are measured when the overall AM pulse height for the XX-gate is tuned to achieve maximal entanglement such that the even-parity two-qubit states, $P_{00}$ and $P_{11}$, are equal ($P_{00}=P_{11}$). The off-diagonal elements are obtained from the amplitude $\Phi$ of a parity oscillation, where the parity is given by $P_{00}+P_{11}-P_{01}-P_{10}$ ($P_{01}$ and $P_{10}$ are the populations of the odd parity two-qubit states). The fidelity can then be calculated as $F =\frac{1}{2}$($P_{00}+P_{11}+\Phi$).\cite{Ballance2016} We use maximum likelihood estimation on experimentally observed data to extract the parameters of the fidelity expression.\cite{Ballance2016} We have performed this analysis for all 55 pairs of qubits in the 11-qubit chain (\fig{XXgates}c) and measure an average fidelity of 97.5$\%$ with a minimum and maximum fidelity of 95.1$^{+0.5}_{-0.7}\%$ and 98.9$^{+0.1}_{-0.3}\%$, respectively. The uncertainty here is determined by a statistical confidence interval on the maximum likelihood estimation. The reported fidelity represents a lower bound of the Bell state creation as we do not correct for SPAM errors on the two-qubit states or errors in single-qubit rotations used to observe the parity oscillations of the Bell state, which on average are 0.7$\%$ and 0.5$\%$ respectively. 

To benchmark our system, we implement two well-known algorithms: the Bernstein-Vazirani (BV) and Hidden Shift (HS). Both of these algorithms have previously been run on trapped-ion\cite{Debnath2016,Fallek2016,Linke2017} and superconducting\cite{Linke2017,Roy2018,Murali2019} systems of up to 5 qubits. By comparing the results of this algorithm to the ideal result, we obtain a direct measure of the system performance, which accounts for our native gates, connectivity, coherence times, gate duration, and all other isolated metrics of system performance. These results can be used as part of a suite of algorithms to compare our hardware with other systems. The qubit number in these results is higher than any comparable published BV or HS results using a programmable quantum computer.\cite{Debnath2016,Fallek2016,Linke2017,Roy2018,Murali2019}

The BV algorithm is an oracle problem in which the user tries to determine an unknown bit string $c$ of size $N$, implemented by a specific oracle. The algorithm takes a binary input string $x$ and performs a controlled inversion of an ancillary bit or qubit based on the bit-wise product of the input and the unknown bit string $c$ modulo two, $f(x) = c\cdot x\text{ (mod 2)}$.\cite{Bernstein1997} For a quantum BV implementation (example shown in \fig{BV}a), a single quantum query is sufficient to determine the bit string $c$.\cite{Grover1997} This is a linear improvement over the best classical algorithm, which requires $N$ queries. The BV algorithm was developed to help separate a class of problems that can be solved in polynomial time on a quantum computer with bounded errors (BQP) from its classical counterpart. For a problem to belong to BQP, a single quantum query success probability needs to be greater than 2/3.\cite{Bernstein1997}
\begin{figure*}[ht]
\includegraphics[width=\textwidth]{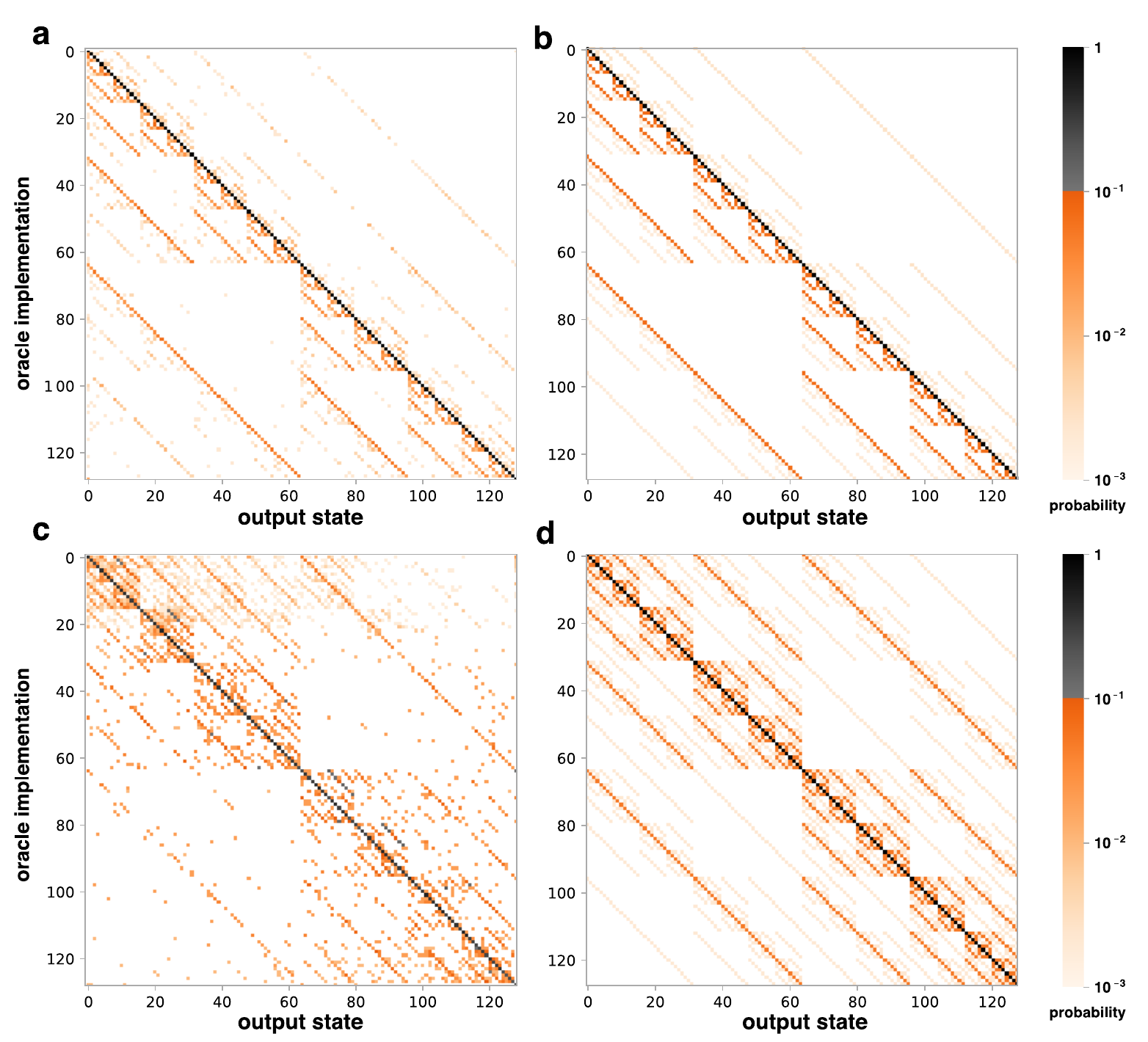}
\caption{Comparison of data to noise model. (a,c) show, respectively, the first 128 by 128 section of Bernstein-Vazirani (BV) and Hidden Shift (HS) results for comparison with the result of a minimal error model including bit-flip errors and detection mis-identification (b,d). Crosstalk errors produce different patterns of errors between the BV and HS due to the structure of the two algorithms.}
\label{fig:errormodel}
\end{figure*}

We compile the BV algorithm into our native gate set, comprised of single-qubit rotations and two-qubit XX-gates. Optimization during compilation reduces the number of needed gates compared to naively translating the textbook circuit from CNOT gates into rotations and XX-gates. The compilation exploits the full connectivity of our qubits, since we do not need SWAP operations. The implementation of BV requires a single-qubit ancilla and a register of $N$ qubits. There are $2^N$ possible bit strings, therefore for our 10-qubit register there are 1024 possible oracle implementations. We measured each implementation 500 times, conditioned upon on the measured ancilla state, and plot the output distribution in \fig{BV}b. Each oracle implementation has, depending on the unknown bit string $c$, between zero and ten two-qubit gates between the ancilla and the qubit register, corresponding to the number of ones in the binary representation of the unknown bit string. The process matrix that maps the encoded oracle to the measured output state is nearly diagonal, resulting in a highly-peaked distribution at the encoded oracle. For our system, the average overlap between output state and unknown bit string is 78$\%$ (\fig{BV}c), where 87.8$\%$ of oracle implementations achieve the 2/3 success criteria defined by BQP. Conditioning the output on the ancilla state results in a 5.1 percentage point increase in the raw success probability of 73$\%$ to 78$\%$ and an 14.5 percentage point increase in the fraction of oracle implementations above the BQP threshold from 73.3$\%$ to 87.8$\%$. The average overlap in \fig{BV}c decreases with the number of two-qubit gates needed in the oracle. The off-diagonal components of the process matrix show errors since these should all have zero population. In \fig{BV}b, the dominant error is single-qubit bit-flips from $\ket{1}$ to $\ket{0}$ during the measurement process, which appear as faint diagonals in the lower left quadrant of the figure (see Methods). However, even for the oracle implementation where we have the lowest success probability, the next-most-probable state  is still 4 times less likely than the correct string.

The HS algorithm consists of two N-bit to N-bit function oracles f and g, which are the same up to a shift by a hidden bit string $s$, such that $g(x) = f(x+s)$. The goal is to determine the hidden shift $s$ by querying the oracles. In our implementation \cite{Rotteler2010} of the HS algorithm, the oracles are inner product or ``bent'' functions $f = \sum_i x_{2i-1} x_{2i}$ and $g = f(x+s)$, where $x$ is the input and $x_i$ is the $i$-th bit of $x$ (an example is shown in (\fig{HS}a). Classically it can be shown that determining the shift $s$ requires $\sqrt{2^N}$ queries where $N$ is the length of the bit string $s$. On a quantum computer, in principle, the shift can be read out in a single query.\cite{vanDam2003,Rotteler2010} In contrast to the BV algorithm, the quantum implementation of the HS algorithm shows an exponential reduction in the number of queries to the oracle compared to a classical computer.\cite{Rotteler2010} 

As with the BV algorithm, we compile the HS algorithm into our native gates. There are $2^N$ available oracle implementations corresponding to the $2^N$ possible hidden bit strings $s$. We execute all 1024 possible implementations on our 10-qubit register (\fig{HS}b). The correct output state is the state corresponding to the hidden shift. The average overlap between the output state and $s$ was 35$\%$ (\fig{HS}c), and of the 1024 oracles, 1017 had most likely output states corresponding to the shift. The success probability for HS is lower and more uniform than that of BV because all of the oracles have the same number of two-qubit gates (10) and many more single-qubit gates (25-40). Every oracle implementation in HS has at least as many gates as the most challenging BV oracle implementation and therefore is more difficult. Given our average single- and two-qubit fidelities, we would not expect to surpass the BQP threshold for the HS oracles. However, the successful determination of the shift was achieved much more frequently than if we sampled a classical distribution where the success probability would have been 0.1$\%$.

In summary, we have constructed the largest programmable quantum computer to date that is capable of performing algorithms. We have used a trapped ion quantum computer to perform the largest quantum implementations of the Bernstein-Vazirani and Hidden Shift algorithms. Using a 10-qubit register, we implement all 1024 possible oracles for each algorithm. We exceed the BQP threshold for 87.8$\%$ of the oracle implementations in the Bernstein-Vazirani algorithm, an application designed to define this complexity class. We also demonstrate 35$\%$ overlap between the measured and expected output states in the implementation of the Hidden Shift algorithm, which is a more demanding application due to its higher gate count and exponential speed up over its classical analog.  The success of both algorithms is a result of high-fidelity native gates and efficient gate compilation and compression in the fully-connected ion trap system. The demonstration of these two canonical algorithms is a starting point for benchmarking any quantum computer. Computing real problems on larger systems with more qubits will require even more gates in the future with even higher quality, and similar standard algorithms to those demonstrated here will likely play a crucial role in benchmarking quantum computers in the future.
\section*{Methods}
\label{sec:Methods}
The dominant error in our data is single-qubit bit flips driven by addressing crosstalk. In the BV algorithm, these errors show up predominantly as qubits that are in state $\ket{0}$ when they should be in state $\ket{1}$ in the output state. We model this as 3$\%$ probability of incorrectly preparing one qubit in the $\ket{0}$ state, which we apply twice: first to model errors that occur in the implementation of the oracle, and then to model errors in the single-qubit gates that occur before readout. We estimate the 3$\%$ probability of incorrectly preparing one qubit by fitting by eye the relative amplitude of error states. Finally, detection mis-identification (measuring the state of one ion as other than it should be) is modeled at 0.2$\%$ probability on the resulting state, the measured SPAM infidelity measured with microwaves (\tab{1Q}). $1/16$ of our measurement results for the BV algorithm are shown in \fig{errormodel}a for comparison with the result of this error model (\fig{errormodel}b).

In the HS algorithm, the crosstalk errors show up as qubits that are measured to be in the wrong state (either $\ket{0}$ or $\ket{1}$). We model this as a 1$\%$ error of incorrectly preparing one qubit in either state, which we apply 5 times, followed 0.2$\%$ detection mis-identification. $1/16$ of our measurement results for the the HS algorithm are shown in \fig{errormodel}c for comparison with the result of this error model (\fig{errormodel}d).

\bibliography{reference}



\section*{Acknowledgements}
The authors would like to thank David Moehring for his guidance in the construction of this apparatus, and the EURIQA team at the University of Maryland and Duke University for sharing their designs and for useful conversations.

\section*{Author Contributions}
Experimental data collected and analyzed by K.W., K.M.B and S.D.; circuit compilation and gate design by Y.N., S.D. and N.G.; the apparatus was designed and built by K.W., K.M.B, J.M.A., S.D., JS.C., N.C.P., M.C., C.C., K.M.H, J.M., J.D.WC., S.A., J.A., P.S., M.W., A.D., A.B., S.M.K., V.C., M.K., C.M. and J.K.; K.M.B. and K.W. prepared the manuscript, with input from all authors.

\section*{Competing Interests}
The authors declare no competing interests.
\section*{Correspondence} 
Correspondence and requests for materials should be addressed to Kenneth Wright~(email: wright@ionq.co).
\begin{table}[ht]
    \centering
    \begin{tabular}{c|c|c|c}
    Ion &  Gate Fidelity    &   SPAM from RB &   SPAM from Microwave \\
    \hline
    0 & 99.57(5) & 99.31(9) & 99.82(4) \\
    1 & 99.62(6) & 99.1(1) & 99.77(5) \\
    2 & 99.18(7) & 99.3(1) & 99.78(5) \\
    3 & 99.25(9) & 99.6(2) & 99.78(5) \\
    4 & 99.40(9) & 99.3(2) & 99.84(4) \\ 
    5 & 99.46(3) & 99.32(7) & 99.77(5)\\
    6 & 99.48(3) & 99.27(6) & 99.82(4)\\
    7 & 99.55(4) & 99.40(8) & 99.83(4)\\
    8 & 99.59(3) & 98.94(6) & 99.80(4)\\
    9 & 99.64(2) & 99.35(4) & 99.79(5)\\
    10 & 99.32(6) & 99.3(1) & 99.79(5)\\
    \end{tabular}
    \caption{Single-qubit  randomized benchmarking (RB) results and microwave SPAM results expressed in percentage ($\%$). To determine single-qubit fidelities for each qubit we apply laser pulses to perform randomized benchmarking for $\pi/2$ gates, using $\pi$ gates to randomize the computational axes. The data are fit to a power law as described in the text. The average single-qubit fidelity is 99.5$\%$. We can obtain SPAM errors from either the RB results or from a microwave pulse. For microwaves we tune the frequency of the microwave to the qubit splitting and the pulse time is set to drive a spin flip from $\ket{0}\rightarrow\ket{1}$, where the fidelity of detecting the $\ket{1}$ state is the measured SPAM fidelity. The average SPAM fidelity from RB is 99.3$\%$ and with microwave based operations the average SPAM fidelity is 99.80$\%$. The uncertainties for the RB results are errors from the fit to a power law (see text) and the uncertainties for microwaves are statistical errors on a binomial distribution, $\sqrt{\frac{P_{\ket{1}}(1-P_{\ket{1}})}{n_{expt}}}$, set by the photon counting statistics.}
    \label{tab:1Q}
\end{table}

\begin{table*}[ht]
    \centering
    \begingroup
    \renewcommand{\arraystretch}{1.5} 
    \begin{tabular}{c|c|c|c|c|c|c|c|c|c||c}
    1	&	2	&	3	&	4	&	5	&	6	&	7	&	8	&	9	&	10	&	\diagbox[dir=SW]{Ion 0}{Ion 1}\\
    \hline
    \hline
98.5$_{-0.3}^{+0.1}$	&	97.7$_{-0.5}^{+0.4}$	&	98.5$_{-0.3}^{+0.1}$	&	97.2$_{-0.5}^{+0.4}$	&	98.5$_{-0.3}^{+0.1}$	&	96.9$_{-0.5}^{+0.5}$	&	97.2$_{-0.5}^{+0.3}$	&	98.7$_{-0.5}^{+0.4}$	&	95.5$_{-0.6}^{+0.4}$	&	97.1$_{-0.3}^{+0.1}$	&	0\\
	&	97.7$_{-0.6}^{+0.4}$	&	98.9$_{-0.3}^{+0.1}$	&	98.2$_{-0.3}^{+0.1}$	&	97.4$_{-0.3}^{+0.1}$	&	97.8$_{-0.3}^{+0.1}$	&	98.1$_{-0.3}^{+0.1}$	&	98.4$_{-0.3}^{+0.1}$	&	97.7$_{-0.5}^{+0.3}$	&	97.9$_{-0.3}^{+0.1}$	&	1\\
	&		&	98.0$_{-0.3}^{+0.2}$	&	97.5$_{-0.4}^{+0.3}$	&	96.5$_{-0.6}^{+0.5}$	&	98.4$_{-0.3}^{+0.1}$	&	98.0$_{-0.3}^{+0.1}$	&	97.2$_{-0.5}^{+0.3}$	&	97.3$_{-0.3}^{+0.1}$	&	96.0$_{-0.6}^{+0.6}$	&	2\\
	&		&		&	96.4$_{-0.5}^{+0.4}$	&	97.4$_{-0.3}^{+0.1}$	&	97.1$_{-0.5}^{+0.4}$	&	98.9$_{-0.3}^{+0.1}$	&	96.0$_{-0.5}^{+0.3}$	&	98.0$_{-0.3}^{+0.1}$	&	97.7$_{-0.3}^{+0.1}$	&	3\\
	&		&		&		&	98.6$_{-0.6}^{+0.3}$	&	97.3$_{-0.4}^{+0.4}$	&	97.3$_{-0.5}^{+0.5}$	&	98.3$_{-0.5}^{+0.4}$	&	97.8$_{-0.3}^{+0.1}$	&	96.5$_{-0.6}^{+0.5}$	&	4\\
	&		&		&		&		&	96.5$_{-0.6}^{+0.4}$	&	97.1$_{-0.5}^{+0.3}$	&	98.4$_{-0.4}^{+0.3}$	&	95.1$_{-0.7}^{+0.5}$	&	96.7$_{-0.6}^{+0.5}$	&	5\\
	&		&		&		&		&		&	96.2$_{-0.6}^{+0.4}$	&	97.2$_{-0.6}^{+0.3}$	&	98.1$_{-0.5}^{+0.4}$	&	98.2$_{-0.5}^{+0.4}$	&	6\\
	&		&		&		&		&		&		&	97.3$_{-0.6}^{+0.4}$	&	98.5$_{-0.3}^{+0.3}$	&	97.3$_{-0.6}^{+0.4}$	&	7\\
	&		&		&		&		&		&		&		&	96.7$_{-0.5}^{+0.4}$	&	97.0$_{-0.6}^{+0.3}$	&	8\\
	&		&		&		&		&		&		&		&		&	97.5$_{-0.5}^{+0.4}$	&	9
    \end{tabular}
    \endgroup
    \caption{Raw fidelity of native two-qubit gates expressed in percentage ($\%$). For each qubit pair, we perform the gate and measure the joint populations of pair qubits as a function of analysis pulse phase angle to determine the parity contrast of the created Bell state. The resulting parity and joint-population are determined using maximum likelihood estimation to extract the fidelities enumerated above. The uncertainties are the 1$\sigma$ confidence interval determined from maximum likelihood estimation. The average fidelity is 97.5$\%$ with a minimum and maximum fidelity of 95.1$\%$ and 98.9$\%$ respectively. }
    \label{tab:XX}
\end{table*}

\end{document}